\documentclass[twocolumn,prb,floatfix,superscriptaddress]{revtex4-1}
\usepackage{graphicx}    
\usepackage{dcolumn}     
\usepackage{bm}          
\usepackage{amsmath}

\begin{document}
	\title{Possible high-$T_c$ superconductivity due to incipient narrow bands originating from hidden ladders in Ruddlesden-Popper compounds}
	\author{Daisuke Ogura}
	\affiliation{Department of Physics, 
		Osaka University, Machikaneyama, 
		Toyonaka, Osaka 560-0043, Japan}
	\author{Hideo Aoki}
	\altaffiliation{Now also at Department of Physics,
		ETH Z\"{u}rich, 8093 Z\"{u}rich, Switzerland}
	\affiliation{Department of Physics, 
		University of Tokyo, Hongo,
		Tokyo 113-0033, Japan}
	\affiliation{Electronics and Photonics Research Institute,
		Advanced Industrial Science and Technology (AIST), Tsukuba, Ibaraki 305-8568, Japan}
	\author{Kazuhiko Kuroki}
	\affiliation{Department of Physics, 
		Osaka University, Machikaneyama, 
		Toyonaka, Osaka 560-0043, Japan}
	\begin{abstract}
		We introduce the concept of ``hidden ladders" for the bilayer Ruddlesden-Popper type compounds: While the crystal structure is bilayer, $d_{xz}$ ($d_{yz}$) orbitals in the relevant $t_{2g}$ sector of the transition metal form a two-leg ladder along $x$ ($y$), since the $d_{xz}$ ($d_{yz}$) electrons primarily hop in the leg ($x,y$) direction along with the rung ($z$) direction.
		This leads us to propose that Sr$_3$Mo$_2$O$_7$ and Sr$_3$Cr$_2$O$_7$ are candidates for the hidden-ladder material, with the right position of $E_F$ from a first-principles band calculation.  
		Based on the analysis of Eliashberg equation, we predict the possible occurrence of high temperature superconductivity in these non-copper materials arising from the interband pair-scattering processes between a wide band and an ``incipient" narrow band on the ladder.
	\end{abstract}
	
	\maketitle
	\section{Introduction}

	If one wants to theoretically search for high-temperature superconductors, 
	which is one of the most challenging problems in condensed matter physics, a guiding principle is imperative. 
	While the large energy scale $t$ of electrons naively favors high-$T_c$ in electron mechanisms of superconductivity, 
	we usually end up with $T_c$ as low as $\sim t/100$, which is still high with the energy scale $t \sim \mathcal{O}(1)$ eV, as in the cuprates, a typical class of unconventional superconductors.  
	On the other hand, $T_c$ of the conventional phonon-mediated superconductors is about several percent of the energy scale of phonons.
	Given this, a real question is how we can design electron-mechanism superconductivity where the ``low" values of $T_c$ could be enhanced.  
	A large potential may be expected, since we have various degrees of freedom from lattice structures, orbital characters, etc, for manipulating electronic pairing mechanisms.
	What would then be the main bottlenecks for enhancing $T_c$ in the unconventional superconductors?
	
	One dilemma is the difficulty in simultaneously realizing a strong pairing interaction and a light renormalized electron mass: 
	A strong electron correlation can mediate strong pairing interactions, but usually makes the electron mass heavy due to the strong quasi-particle renormalization.  
	For instance, if we consider the spin-fluctuation-mediated pairing in the single-band Hubbard model on a square lattice, which is the simplest model for the high-$T_c$ cuprates, we have indeed the strong pairing interaction simultaneously with the strong quasiparticle renormalization. 
	The former works in favor of superconductivity, whereas the latter works against it.
	
	In a previous paper\cite{Kuroki2005_wide-narrow}, one of the present authors and his coworkers proposed a possible way to circumvent this problem.
	Let us consider a system where narrow and wide bands coexist due to a lattice structure such as a ladder, and set the Fermi level close to, but not right within, the narrow band.
	Then the electrons in the wide band, which are not very strongly renormalized, can form Cooper pairs with a pairing interaction strongly mediated by the large number of interband scattering channels, where the narrow (or even flat) band acts as intermediate states while its large density of states does not cause a heavy mass for the electrons that form the Cooper pair.
	In fact, pairing interaction originating from a narrow band lying below, but not far away from, the Fermi level has attracted much attention recently in the context of the ``incipient band" in iron-based superconductors\cite{Hirschfeld2011_incipient, Miao2015_ironbased111FS, Wang2011_122FRG, Bang2014_122shadowgap, Chen2015_incipient, Bang2016_dynamicaltuning, Mishra_2016} especially after the observation of missing hole Fermi pockets in some of the iron-based compounds\cite{Guo2010_ironbased122, Qian2011_ironbased122FS, Wang2012_ironbased11-STO, Tan2013_ironbased11-STO, Miao2015_ironbased111FS, Niu2015_ironbased11, Sunagawa2016_ironbased122FS}.  The idea for narrow bands in Ref.\cite{Kuroki2005_wide-narrow} may be regarded as a precursor of the incipient band in iron-based superconductors.  
	Superconductivity involving flat bands has also been discussed recently in the context of the flat-band superconductivity\cite{Kobayashi2016_flatband}, where the virtual pair scattering between the flat and dispersive bands is also at work.
	The incipient-band-induced superconductivity exploits finite-energy spin fluctuations, whose importance has recently been noted in other models as well\cite{Nakata2017_finite-energy}.
	
	\begin{figure*}[t]
	\centering
	\includegraphics[width=12cm,clip]{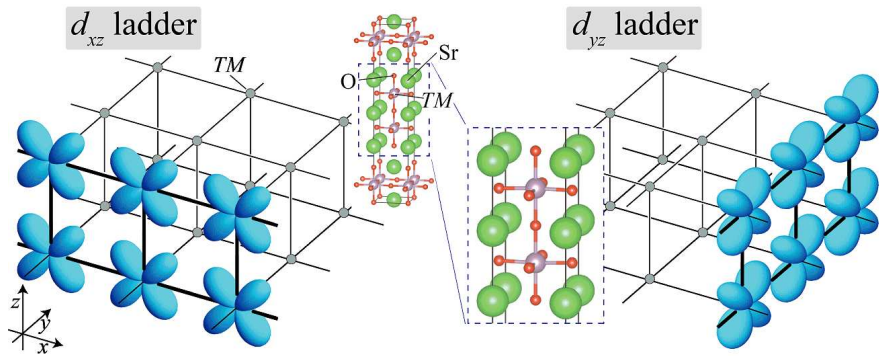}
	\caption{
		Schematics of ``hidden ladders" composed of the $d_{xz}$ (left panel) and $d_{yz}$ (right) orbitals in the bilayer Ruddlesden-Popper compounds Sr$_3${\it TM}$_2$O$_7$ ({\it TM} indicates a transition metal). The middle panel depicts the crystal structure.
	}
	\label{lattice}
	\end{figure*}
	In Ref.\cite{Kuroki2005_wide-narrow}, a two-leg Hubbard ladder was considered as a candidate model in which the above situation can be realized. In fact, the study of unconventional superconductivity in ladder-type materials has a long history, starting from the seminal proposal by Dagotto and Rice\cite{Dagotto1992_ladder,Rice1993_ladder,Dagotto-Rice1996_ladder}. Experimentally, superconductivity with a $T_c$ exceeding 10 K was then observed in (Sr,Ca)$_{14}$Cu$_{24}$O$_{41}$\cite{Nagata1998_14-24-41}.
	The subsequent proposal in Ref.\cite{Kuroki2005_wide-narrow} was that a much higher $T_c$ may take place in two-leg cuprate ladder compounds, provided that $\sim 30$ \% of electron doping is achieved.
	Namely, in a single-orbital Hubbard model on a two-leg ladder with only the nearest-neighbor hopping, there are two (bonding and antibonding) bands with an identical shape separated in energy by $2t_\mathrm{rung}$, where $t_\mathrm{rung}$ is the hopping in the rung direction.
	When the second-neighbor (diagonal) hoppings are introduced, one of the bands becomes narrower, while the other becomes wider. 
	For the sign of the diagonal hopping corresponding to the real material, a large electron doping is required to put the Fermi level just above the top of the narrow band. 
	A snag, however, is that doping carriers into the cuprate ladders is known to be difficult\cite{Kojima2001_ladder-review}; especially electron doping with such a large amount may be rather unrealistic.
	
	In the present paper, we propose an entirely different, and realistic, way to realize the above-mentioned situation where narrow and wide bands coexist with the Fermi level lying close to the edge of the narrow one.
	We propose to start with the Ruddlesden-Popper-type layered perovskites with two layers in a unit cell.
	These bilayer compounds may first seem to have nothing to do with ladders.
	However, if we consider a case in which the $t_{2g}$ orbitals are relevant (i.e., forming the bands crossing $E_F$), an electron in, say, the $d_{xz}$ orbital selectively hops in the $x$ direction as well as in the $z$ direction normal to the bilayer.
	This means that the $d_{xz}$ orbitals form a ladder with the $x$ and $z$ directions being the leg and rung directions, respectively.
	Similarly, $d_{yz}$ orbitals form a ladder in the $y$ and $z$ directions, as schematically shown in Fig.\ref{lattice}.  
	Hence ladder-like electronic structures are, in fact, {\it hidden} in the apparently bilayer Ruddlesden-Popper materials.  
	To be precise, the $d_{xz}$ and $d_{yz}$ orbitals are weakly hybridized with each other, and also with the $d_{xy}$ orbital, so that the quasi-one dimensional ladder character is slightly degraded, but they should still essentially be ladder bands.
	
	\section{Formulation}
	\subsection{Intuitive idea}
	As mentioned in the Introduction, the band filling is an essential factor if we want to have the narrow band as incipient in the ladder-like electronic structure to realize a high-$T_c$ superconductivity.  
	 An electron doping of $\sim 30$ \% is required\cite{Kuroki2005_wide-narrow} for the cuprate ladder as mentioned, where the $d_{x^2-y^2}$ (i.e., an $e_g$) orbital is relevant. 
	By contrast, if the $d_{xz}$ and $d_{yz}$ (i.e., $t_{2g}$) orbitals are relevant, the sign of the hopping integral is reversed from the case of $d_{x^2-y^2}$, so that the ideal situation should be $\sim 30$ \% {\it hole} doping, which corresponds to $1/3$ filling, namely, two electrons on the three $t_{2g}$ orbitals.  
	If we consider non-copper materials of the form Sr$_3${\it TM}$_2$O$_7$ ({\it TM} indicates a transition metal), the valence of the transition metal should be +4. This implies that if we want to have a $d^2$ electron configuration, we can take, instead of the hole doping, stoichiometries Sr$_3$Mo$_2$O$_7$ (with a 4$d$ transition metal Mo) and Sr$_3$Cr$_2$O$_7$ (with a 3$d$ transition metal Cr) for possible candidates as the hidden ladder materials.
	
	\subsection{Band calculation and many-body analysis}
	
	We start with a first-principles band calculation for these materials with the WIEN2k package\cite{Blaha2001wien2k}, assuming the ideal case of no long-range orders and adopting the experimental lattice structures in Refs.\cite{Steiner2004_Mo327str,Castillo2007_Cr327str}.
	In the first-principles calculation, the Perdew-Burke-Ernzerhof parametrization of the generalized gradient approximation\cite{Perdew_PBE1996} was used, with 1000 $k$-meshes and  $RK_{\rm max}=7$.

	To proceed to many-body analysis, we first construct a 	three-dimensional tight-binding model with six orbitals (three $t_{2g}$ orbitals in each layer) for Sr$_3$Mo$_2$O$_7$ constructed from the maximally-localized Wannier orbitals using the Wannier90 code\cite{Mostofi2008wannier90} and the wien2wannier interface\cite{Kunevs2010wien2wannier}.  
	
	To take account of the electron correlation effect beyond the LDA/GGA level for this tight-binding model, we introduce multi-orbital Hubbard-type interactions on each atomic site with the Hamiltonian,
	\begin{eqnarray}
	H_{\mathrm{int}}=\sum_{i}\left(\sum_{\mu}U_{\mu}n_{i\mu\uparrow}n_{i\mu\downarrow}+\sum_{\mu>\nu}\sum_{\sigma\sigma^\prime}U'_{\mu\nu}n_{i\mu\sigma}n_{i\mu\sigma^\prime} \right. \nonumber \\ \left.
	-\sum_{\mu\ne\nu}J_{\mu\nu}\bm{S}_{i\mu}\cdot\bm{S}_{i\nu}+\sum_{\mu\neq\nu}J^\prime_{\mu\nu}c_{i\uparrow}^{\mu\dag}c_{i\downarrow}^{\mu\dag}c_{i\downarrow}^{\nu}c_{i\uparrow}^{\nu}\right), \nonumber
	\end{eqnarray}
	where $i$ labels the site and $\mu,\nu$ are the orbitals in a notation adopted in Ref.\cite{Kuroki2009_Pnictogen-height}.
	To treat the many-body effect, we apply the multi-orbital fluctuation exchange (FLEX) approximation\cite{Bickers1989,Bickers1991}. 
	We set the on-site interactions as $U = 3.0$ eV or $U=2.5$ eV, along with $U^\prime (= U-2J$ for preserving the orbital SU(2) symmetry) and $J = J^\prime = U/10$.
	These are typical values evaluated with the constrained random phase approximation\cite{Vaugier_2012_cRPA_3d-4d, Mravlje_2011_cRPA, Jang_2016_cRPA} or the constrained LDA\cite{Pchelkina_2007_SRO_cLDA} method for the Hubbard-type model for the $d$-orbitals of transition metals.
	The temperature is fixed at $T = 0.01$ eV, and we take $N=32\times 32\times4$ $k$-point meshes and 2048 Matsubara frequencies. 
	
	In the FLEX approximation, the effective electron-electron interaction for obtaining the self-energy is calculated by collecting bubble- and ladder-type diagrams consisting of the renormalized Green's function $G(k)$, namely, by summing up powers of the irreducible susceptibility,
	\begin{eqnarray}
	\chi^{\rm irr}_{l_1l_2, l_3l_4}(q) \equiv -\frac{T}{N}\sum_{k} G_{l_3l_1}(k)G_{l_2l_4}(k+q). \nonumber
	\end{eqnarray}		
	Here $k$ is a shorthand for the wave number and the Matsubara frequency, and $l_i$ denotes orbitals.  
	The renormalized Green's function is obtained by solving the Dyson equation.  
 	We then obtain the singlet pairing interaction $\Gamma(q)$ mediated mainly by spin fluctuations, which is plugged into the linearized Eliashberg equation for superconductivity,
	\begin{eqnarray}
	\lambda \Delta_{ll^{\prime}}(k) &=& -\frac{T}{N}\sum_{k^{\prime}m_i}\Gamma_{lm_1 m_4l^{\prime}}(k-k^{\prime})G_{m_1m_2}(k^{\prime})\nonumber\\
	&&\times \Delta_{m_2m_3}(k^{\prime}) G_{m_4m_3}(-k^{\prime}). \nonumber
	\end{eqnarray}	
	Since the eigenvalue $\lambda$ of the equation reaches unity at $T=T_c$, here we adopt $\lambda$, obtained at a fixed temperature, to measure how close the system is to superconductivity. 	

	\section{Results}	
	We first display the band structures of  for Sr$_3$Mo$_2$O$_7$ and Sr$_3$Cr$_2$O$_7$ obtained from the first-principles band calculation.
	The result in Fig.2 indeed shows that the $d_{xz}$ and $d_{yz}$ orbitals form quasi-one-dimensional narrow (antibonding) and wide (bonding) bands in both materials, as identified from the band dispersion along P-N and N-$\Gamma$.  

	\begin{figure}[h!]
		\centering
		\includegraphics[width=8.5cm,clip]{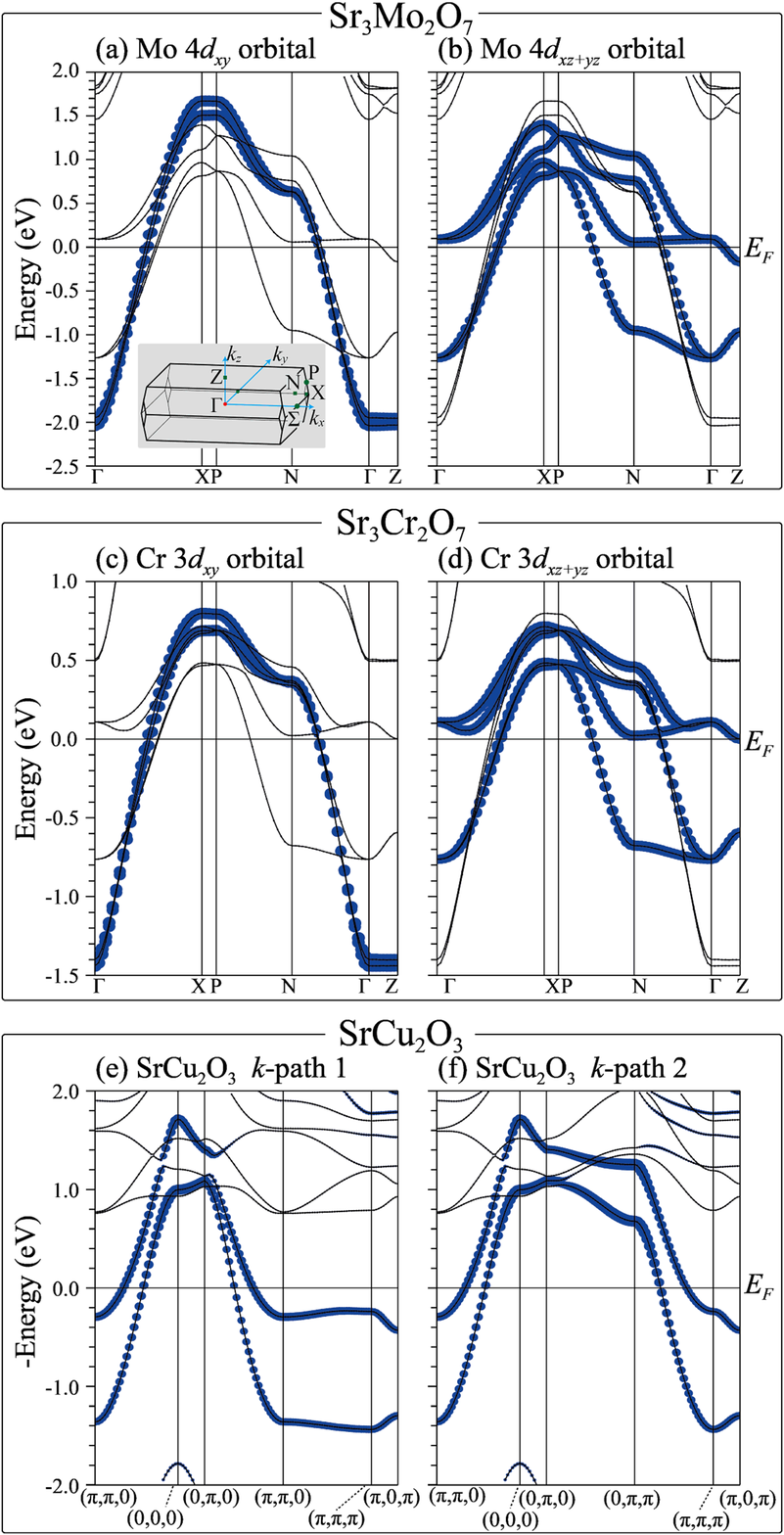}
		\caption{Band structures of (a,b) Sr$_3$Mo$_2$O$_7$ and (c,d) Sr$_3$Cr$_2$O$_7$ obtained from the first-principles calculation. The Brillouin zone is shown in the inset. The thickness of the lines in the left(right) panels depicts the weight of the $d_{xy}$($d_{xz,yz}$) orbital character. (e,f) For comparison, the band structure of a ladder cuprate SrCu$_2$O$_3$ is displayed along two paths in the conventional Brillouin zone, where the thickness of the lines represents the $d_{x^2-y^2}$ orbital character. The point $(\pi,\pi,0)$ in panels (e,f) corresponds to $\Gamma$ point in panels (a-d) because the sign of the hopping is reversed. Note that panels (e) and (f) superimposed form a band structure quite similar to those in (b) and (d), respectively, except for the position of the Fermi level.}
		\label{band}
	\end{figure}

	For comparison, we have also performed a band calculation for the well-known two-leg ladder cuprate SrCu$_2$O$_3$, with the lattice structure given in Ref.\cite{Sparta2006_SrCu2O3,Muller_SrCu2O3}.  
	In the result displayed in bottom panels of Fig.\ref{band} we have reversed the sign of the energy to facilitate comparison, since the roles of electrons and holes are exchanged between the $e_g$ (cuprate) and the $t_{2g}$ (present) systems.  
	We can see that the band structures of the $d_{xz/yz}$ orbitals of the bilayer Ruddlesden-Popper compounds are very similar to that of the ladder compound. However, an important difference lies in the position of the Fermi level: 
	In the Ruddlesden-Popper compounds $E_F$ is situated very close to the bottom edge of the narrow bands, namely, the narrow bands can be regarded as incipient.  This is in sharp contrast to the case of SrCu$_2$O$_3$, where $E_F$ sits far away from the narrow band edge.
	These results confirm the intuitive expectation presented above.

	\begin{figure}[tbp]
		\centering
		\includegraphics[width=5.0cm,clip]{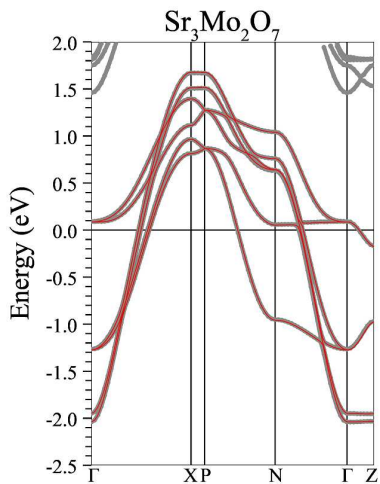}
		\caption{Band structure (red lines) of the six-orbital tight-binding model for Sr$_3$Mo$_2$O$_ 7$ constructed from the maximally-localized Wannier orbitals. Gray lines represent the first-principles calculation.}
		\label{wannier}
	\end{figure}

	In Fig.\ref{wannier}, we superpose the band structure of the six-orbital tight-binding model for Sr$_3$Mo$_2$O$_7$ constructed from the maximally-localized Wannier orbitals on that of the first-principles calculation.
	We can see that the fit is quite accurate for all the bands intersecting the Fermi level.  

\begin{figure}[tbp]
		\centering
		\includegraphics[width=8.0cm,clip]{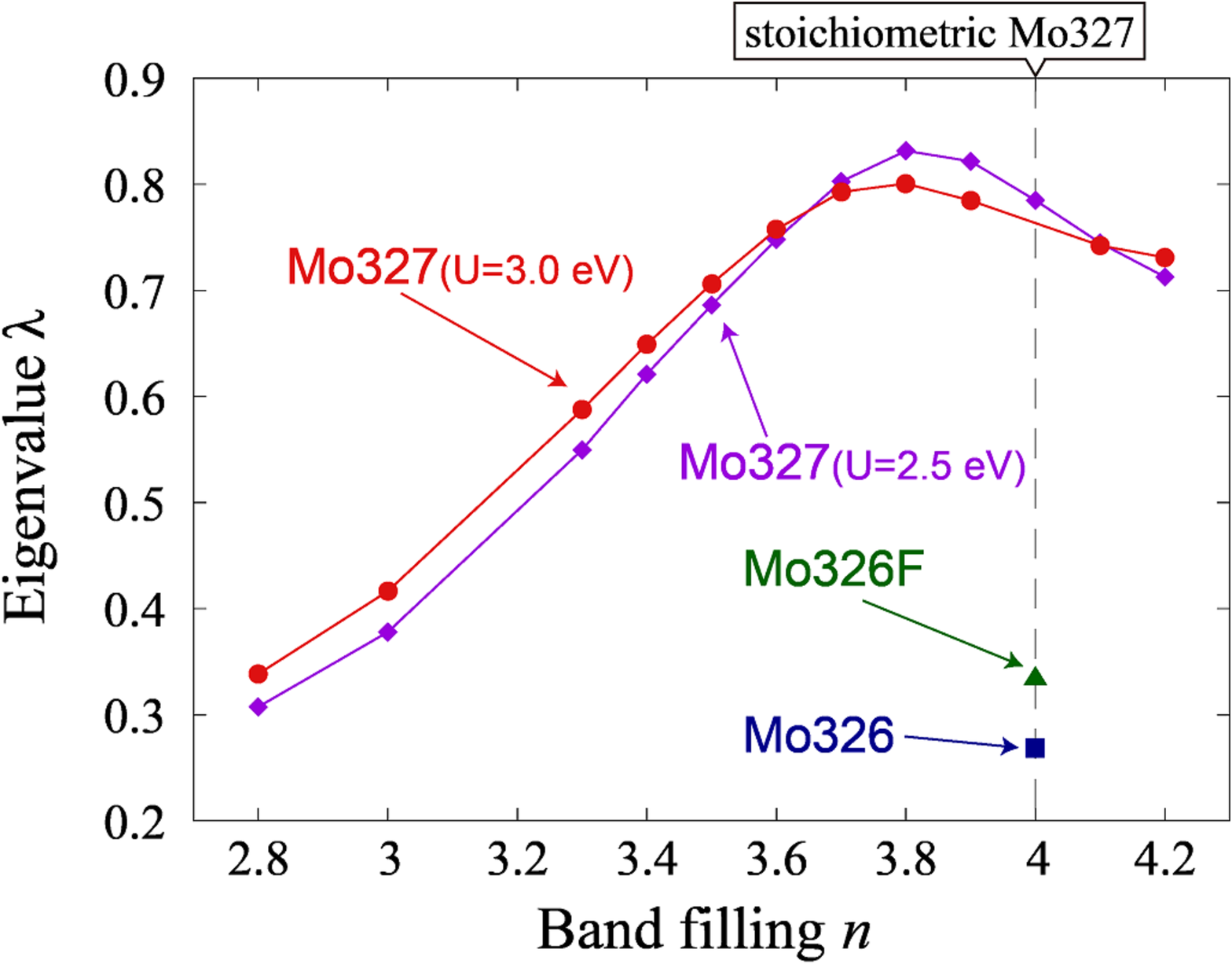}
		\caption{FLEX result for the band filling dependence of the eigenvalue $\lambda$ of the linearized Eliashberg equation for Sr$_3$Mo$_2$O$_ 7$ for $U=3.0$ eV (red dots) and $U=2.5$ eV (purple diamonds).  The vertical dashed line represents the stoichiometric point ($n=4$).  
		Also displayed are the results for the 326 compound Sr$_3$Mo$_2$O$_6$ (blue square) and a F-doped Sr$_3$Mo$_2$O$_6$F (green triangle) at $n=4$.}
		\label{lambda}
	\end{figure}

	Using the six-orbital model, we now study the possibility of superconductivity in Sr$_3$Mo$_2$O$_7$. Figure \ref{lambda} displays the FLEX result for the eigenvalue $\lambda$  of the Eliashberg equation against the band filling. 
	Here the filling $n$ is defined as the average number of electrons per unit cell, so that $n=4$ corresponds to the stoichiometric $d^{2}$ electron configuration. 
	As seen from the result, $\lambda$ is qualitatively unchanged when we change $U=2.5-3.0$ eV within a realistic range, except that for $U=2.5$ eV, (i) the maximum value of $\lambda$ is somewhat enhanced from the case of  $U=3$ eV, and (ii) a decrease of $\lambda$ in the left side of the dome becomes slightly steeper. 
	In both cases, we can see that $\lambda$ attains a maximum $\simeq 0.8$ near the stoichiometric band filling of $n=4$, 	which can be regarded to signal a high-$T_c$ in this material, since the value of $\lambda$ is similar to that of the cuprate superconductor HgBa$_2$CuO$_4$ having $T_c \simeq 90$ K as obtained with FLEX for systematic models that qualitatively explains the material dependence of $T_c$ among the cuprates\cite{Sakakibara2010}.  This is a key result in the present work. 
	
	To be precise, $\lambda$ in Fig.\ref{lambda} is peaked at $n\simeq 3.8$, but decreases as we move away from that point, which can be interpreted as follows:  
	For a too small band filling, superconductivity is suppressed because the narrow band lies too far from the Fermi level.  
	When the band filling is too large, on the other hand, the Fermi level lies deep inside the narrow band, so that the system is put close to antiferromagnetic order, suppressing $\lambda$.
	Hence we conclude that the superconductivity should occur around the stoichiometric point ($n=4$) with an optimized band filling at a slightly hole-doped $n\simeq 3.8$.   
	
	If we turn from the Mo compound to the 3$d$ compound, Sr$_3$Cr$_2$O$_7$, we may also expect high-$T_c$ superconductivity, since its band structure is similar to that of Sr$_3$Mo$_2$O$_7$ as seen in Fig.2.  
	
	\section{Discussion}
	The materials, Sr$_3$Mo$_2$O$_7$ and Sr$_3$Cr$_2$O$_7$, 
	have in fact been synthesized in the past\cite{Kafalas1972_Cr327syn,Kouno2007_Mo327syn,Hosono2015_FIRST}.
	Sr$_3$Cr$_2$O$_7$ turns out to be an antiferromagnetic insulator, while Sr$_3$Mo$_2$O$_7$ a Pauli-paramagnetic metal, where 
	superconductivity has not been observed so far in these materials.  
	Let us then explore the reasons for that.  
	Sr$_3$Cr$_2$O$_7$ has two $d$ electrons within the three $t_{2g}$ orbitals, so that we have 
	a possibility of orbital ordering which will put the system in an insulating state\cite{Jeanneau2017_note}.  
	In order to realize superconductivity, such orders have to be suppressed by, e.g., applying pressure or doping  carriers.  
	If we apply a physical pressure, for instance, the system will become more weakly correlated due to an increased band width.
	Applying chemical pressure with isovalent elemental substitution such as Ca doping to Sr sites can also exert a similar effect.
	As for the carrier concentration, hole doping in the $d_{xz/yz}$ orbitals is desirable for superconductivity in our viewpoint, but this may be chemically difficult to achieve. 
	One way to  attain this is to apply uniaxial pressure along the $c$-axis (in a single crystal), so that the energy level of the $d_{xz/yz}$ orbitals is raised relative to that of $d_{xy}$. 
	Another possible way for carrier doping is to utilize field-effect transistor techniques, which enables fine tuning of the carrier concentration.
	
	In the case of Sr$_3$Mo$_2$O$_7$, we speculate that oxygen deficiencies, especially at the sites between adjacent MoO planes, may be hindering superconductivity.  
	Since the Cooper pair considered here is formed across the two planes (i.e., on each rung of the ladder) in real space, the loss of these oxygens would be detrimental to the superconductivity.  
	Indeed, Ref.\cite{Poltavets2006_Ni326} reports that in La$_3$Ni$_2$O$_7$ having the same bilayer Ruddlesden-Popper structure as Sr$_3$Mo$_2$O$_7$, oxygen vacancies are located solely at the sites between adjacent NiO layers\cite{Pardo2011_Ni327oxygen}.  
	As for Sr$_3$Mo$_2$O$_7$, there is in addition some difficulty in the synthesis of samples with the designated stoichiometric composition\cite{Eisaki_privatecommunication}. 
	In Ref.\cite{Kouno2007_Mo327syn}, there remains ambiguity in the precise chemical composition, since the samples are fabricated with an oxygen buffer, Ti$_2$O$_3$.  
	In high-pressure synthesis methods, the samples tend to be obtained with oxygen contents ($\simeq$O$_{6.3}$) somewhat reduced from the nominal one\cite{Eisaki_privatecommunication}.
	Thus the oxygen deficiencies at the sites between adjacent MoO planes are very likely to be present in the samples fabricated so far.

	To demonstrate that deficiencies of the oxygen atoms connecting the two planes are fatal for superconductivity, we have actually performed a calculation for two hypothetical materials: a ``326" compound Sr$_3$Mo$_2$O$_6$ and a F-doped Sr$_3$Mo$_2$O$_6$F.  
	The crystal structure of Sr$_3$Mo$_2$O$_6$, where the oxygens connecting two MoO layers are totally missing, is determined with the structural optimization calculation using the VASP package\cite{Krasse1993VASP,Krasse1999VASP}. 
	Fluorine-replaced Sr$_3$Mo$_2$O$_6$F, on the other hand, mimics Sr$_3$Mo$_2$O$_{6.5}$.
	We employ the structure average between Sr$_3$Mo$_2$O$_7$ and Sr$_3$Mo$_2$O$_6$ for this hypothetical material.
	As in the case of Sr$_3$Mo$_2$O$_7$, we perform a FLEX calculation for the six-orbital model for each material constructed from the first-principles calculation with the maximally-localized Wannier basis.
	U=3.0 eV is adopted here.
	The obtained eigenvalues superposed in Fig.\ref{lambda} indeed show that the oxygen deficiencies drastically suppress superconductivity. 	
	To show this we have fixed the filling at $n=4$ to single out the effect of degraded ladder-like structure, oxygen deficiencies should introduce some electron doping, which makes the Fermi energy deeper into the narrow band, which may further degrade superconductivity. 
	Thus the direction suggested from these arguments is to synthesize the material with as little oxygen deficiencies as possible. 

	\begin{figure}[h!]
		\centering
		\includegraphics[width=9cm,clip]{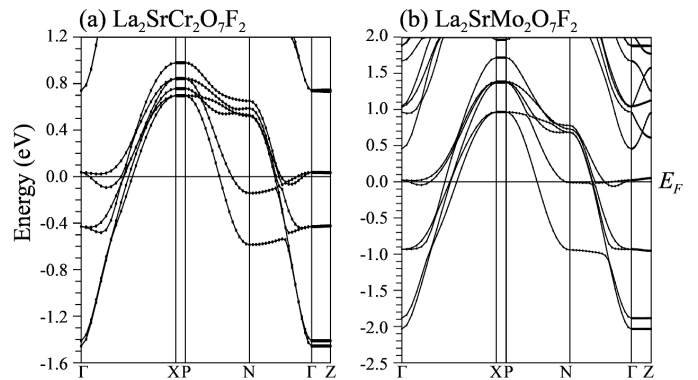}
		\caption{Band structures of fluorine-intercalated compounds (a) La$_2$SrCr$_2$O$_7$F$_2$ and (b) La$_2$SrMo$_2$O$_7$F$_2$.}
		\label{band327F2_vasp}
	\end{figure}
	
	While the present study focuses on bilayer Ruddlesden-Popper materials, the present idea can be extended to some other bilayer materials, such as fluorine-intercalated compounds La$_2$SrCr$_2$O$_7$F$_2$ and La$_2$SrMo$_2$O$_7$F$_2$.
	In these compounds, fluorine atoms are intercalated between the bilayer structures that are the same as those in Sr$_3$Cr$_2$O$_7$ and Sr$_3$Mo$_2$O$_7$, so that the band structure should be quite similar.
	In order to confirm this expectation, we have performed a first-principles band calculation for the ideal case of no long-range orders or lattice distortion.  
	First we determined the crystal structure of La$_2$SrCr$_2$O$_7$F$_2$ and La$_2$SrMo$_2$O$_7$F$_2$ with the VASP package.
	The virtual crystal approximation(VCA) is adopted to take account of the effect of the substitution of La for Sr.
	The band structures obtained with the optimized lattice structures are shown in Fig.\ref{band327F2_vasp}, which are indeed quite similar to those of Sr$_3${\it TM}$_2$O$_7$ ({\it TM} $= {\rm Mo, Cr}$) except that the three dimensionality is somewhat reduced. 
	Given the band structure similar to those of the Ruddlesden-Poppers 327 systems, high-$T_c$ superconductivity is again expected in these systems under certain conditions for carrier doping, etc. In fact, La$_2$SrCr$_2$O$_7$F$_2$ has been synthesized\cite{Zhang2016_La2SrCr2O7F2}, but turns out to be an antiferromagnetic insulator with tilted CrO octahedra.  
	As in Sr$_3$Cr$_2$O$_7$, suppression of the magnetic order (and possibly the lattice distortion) should be necessary for superconductivity.
	
	\begin{figure}[ht]
		\centering
		\includegraphics[width=7.5cm,clip]{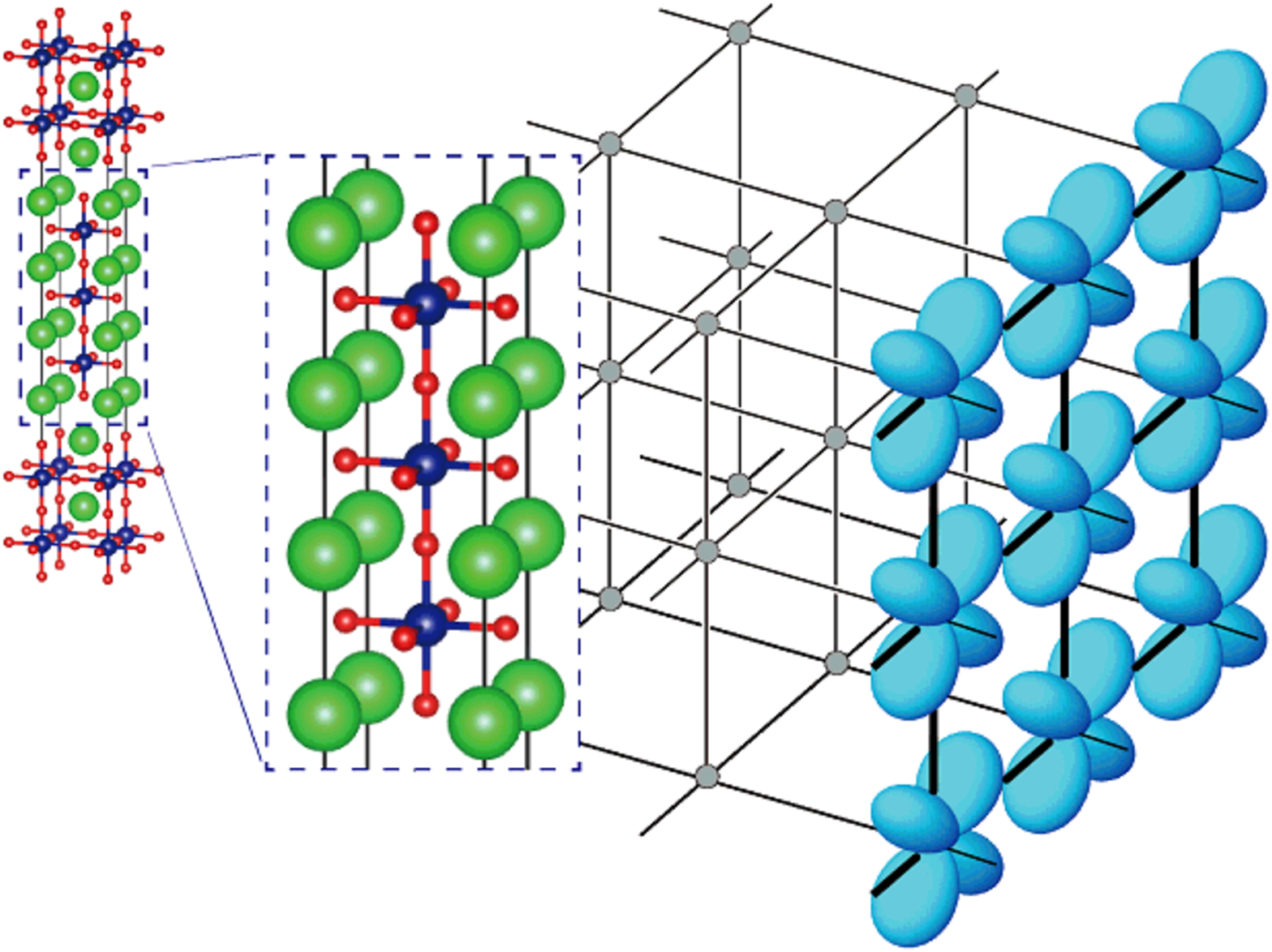}
		\caption{
			A schematic hidden ladder in the triple-layer Ruddlesden-Popper materials, here displayed for $d_{yz}$ orbitals. 
		}
		\label{lattice_tri}
	\end{figure}

	\begin{figure*}[pt]
	\centering
	\includegraphics[width=17cm,clip]{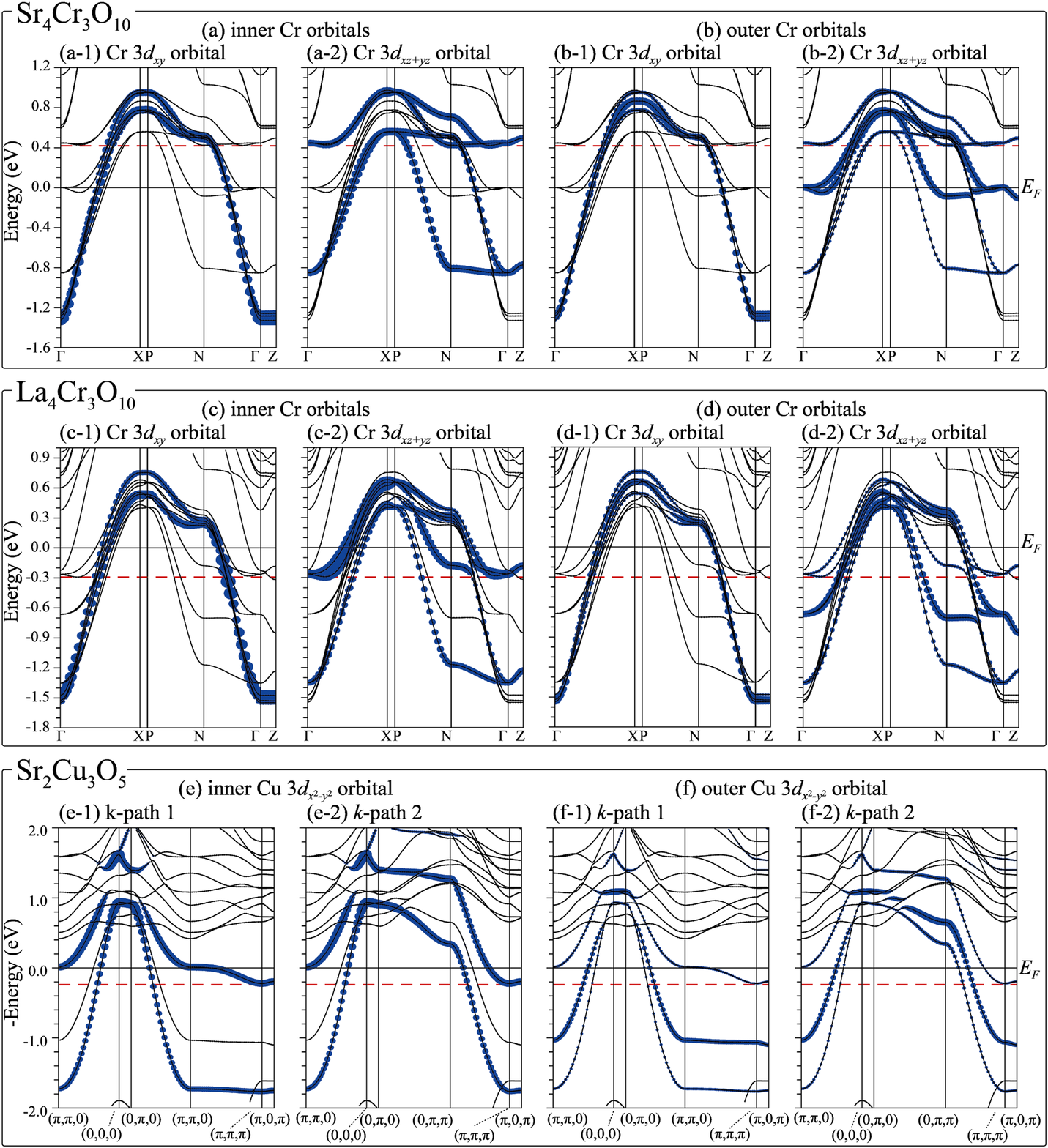}
	\caption{
		Band structures of triple-layer compounds (a, b) Sr$_4$Cr$_3$O$_{10}$ and (c, d) La$_4$Cr$_3$O$_{10}$.
		The thickness of the lines in the left(right) panels depicts the weight of the $d_{xy}$($d_{xz,yz}$) orbital character. 
		(e, f) For comparison, the band structure of a three-leg ladder cuprate Sr$_2$Cu$_3$O$_5$ is shown along two paths in the conventional Brillouin zone, where the thickness of the lines represents the $d_{x^2-y^2}$ character. Note that the sign of the energy is reversed to facilitate comparison between the $t_{2g}$ (present) and $e_g$ (cuprate) system. As in the two-layer case, panels (e-1) and (e-2) [(f-1) and (f-2)] superimposed form a band structure quite similar to those in (a-2) and (c-2) [(b-2) and (d-2)]. Red dashed lines indicate the edge of the narrow bands.
	}
	\label{4310_band}
	\end{figure*}

	We can further extend the present concept for the double-layer compounds to {\it triple}-layer cases, e.g., {\it A}$_4${\it TM}$_3$O$_{10},$ where we can envisage that three-leg-ladder-like electronic structures are hidden.
	Namely, if one adds the third layer in Fig.\ref{lattice} to consider the $d_{xz}$ and $d_{yz}$ orbitals in a similar manner, one can find a pair of three-leg ladders as schematically depicted in Fig.\ref{lattice_tri}.
	In order to show that the above expectation is valid,  we have performed a band calculation for two compounds, Sr$_4$Cr$_3$O$_{10}$ and La$_4$Cr$_3$O$_{10}$. 
	As for Sr$_4$Cr$_3$O$_{10}$, which has actually been synthesized in the past \cite{Kafalas1972_Cr327syn}, we adopt the experimental crystal structure.
	Since La$_4$Cr$_3$O$_{10}$ has not been synthesized so far, we have theoretically determined the crystal structure with the structure optimization.
	The resulting band structures are displayed in Fig.\ref{4310_band}.
	For comparison, we have also performed a band calculation for a three-leg ladder cuprate Sr$_2$Cu$_3$O$_5$, with the lattice structure given in Ref.\cite{Kazakov_1997_3leg-cuprate}.
	The result is shown in the bottom panels of Fig.\ref{4310_band} with the reversed sign of the energy to facilitate comparison as in the two-leg ladder case.
	Since there are two inequivalent [inner- and outer-layer(chain)] Cr(Cu) sites, we show the orbital characters of $t_{2g}$($3d_{x^2-y^2}$) orbitals for each of them. The band structure of the $d_{xz}$ and $d_{yz}$ orbitals is indeed seen to be similar to that of the three-leg ladder.
	Theoretically, it has been known that three-leg ladders should also exhibit superconductivity\cite{Arrigoni1996_3-leg, Schulz1996_ladder, Kimura1996_3leg-ladder, Lin-Balents-Fisher1997_n-leg} as in the two-leg case.  According to our on-going FLEX investigation on the three-leg Hubbard ladder model\cite{Matsumoto_2017_arXiv}, superconductivity is strongly enhanced when the chemical potential is in the vicinity of the edge of the narrow band, which is depicted by red dashed lines in Fig.\ref{4310_band}, but the three-leg ladder cuprate Sr$_2$Cu$_3$O$_5$ itself is notorious for being unable to dope carriers. 
	Observing the band calculation results of the triple-layer materials, we may expect high-$T_c$ superconductivity if (La, Sr)$_4$Cr$_3$O$_{10}$ can be synthesized with a certain content ratio of La and Sr close to unity.

	\section{Conclusion}
	To summarize, we have introduced the concept of the ``hidden ladder" electronic structure in the Ruddlesden-Popper layered perovskites, where anisotropic $d$ orbitals of the transition metal give rise to inherent ladder-like electronic structures.  
	We have proposed actual candidates in Sr$_3$Mo$_2$O$_7$ and Sr$_3$Cr$_2$O$_7$ from first-principles band calculations.  
	We have then pointed out that coexisting narrow and wide bands arising from the ladder can host high-temperature superconductivity due to the strong pairing interaction originating from the virtual pair-hopping processes between the wide and incipient narrow bands, 
	which is supported by a FLEX calculation for Sr$_3$Mo$_2$O$_7$, where no or only slight hole doping is required. The idea can be further extended to fluorine intercalated and triple-layer compounds.
	To realize superconductivity, we note that the possibility of Mott transition as well as occurrence of other orders such as orbital ordering have to be considered, details of which will be an important future work.
	
	We appreciate Hiroshi Eisaki, Kenji Kawashima, Shigeyuki Ishida, Hiraku Ogino, and Yoshiyuki Yoshida for illuminating discussions on experimental factors in realizing the hidden ladders in actual materials.  
	We also thank Masayuki Ochi and Hidetomo Usui for fruitful discussions, 
	and Katsuhiro Suzuki for assistance with the multi-orbital FLEX code. 
	Part of the numerical calculations were performed at the Supercomputer Center,
	Institute for Solid State Physics, University of Tokyo.  
	In Figs.1 and 6, the VESTA software\cite{Momma_2011_vesta} is used to plot the crystal structure.  
	This study has been supported by JSPS KAKENHI (Grants JP26247057 and JP16H04338), a Grant-in-Aid for Scientific Research on Innovative Areas (JP17H05481), and H.A. by ImPACT Program of Council 
	for Science, Technology and Innovation, Cabinet Office,
	Government of Japan (2015-PM12-05-01) from JST.  
	D.O. acknowledges a support from a Grant-in-Aid for JSPS Research Fellow Grant No. 16J01021.
	%
	%
	
	%
	
	%
	
	%
	\bibliography{ronbun}
	
\end{document}